\documentclass[10pt]{ijiss}
\def\currentvolume{?}
\def\currentissue{?}
\def\currentyear{2007}
\def\month{March}
\pagespan{1}{13}
\copyrightinfo{2007}{} 

\begin{document}
\newtheorem{Th}{Theorem}
\newtheorem{Lemm}{Lemma}
\newtheorem{Cor}{Corollary}
\newtheorem{Prop}{Proposition}
\newtheorem{Rem}{Remark}

\def\cal{\mathcal}
\def\pf{{\noindent \em Proof. \hspace{.01in}}}
\newcommand{\eop}{\hfill \vrule height 7pt width 4pt depth 0pt}

\title[Non-Uniform Spin Chains]
{Boundary Effects in Non-Uniform Spin Chains}

\author[K.~E.~Feldman]{K.~E.~Feldman}
\address{
  DPMMS, CMS, University of Cambridge,
  Wilberforce Road, Cambridge,
  CB3 0WB, UK
}
\email{k.feldman@dpmms.cam.ac.uk}
\urladdr{http://www.dpmms.cam.ac.uk/$\sim$kf262/}



\date{January 10, 2007 and, in revised form, March 07, 2007.}

\thanks{This research was supported by EPSRC grant GR/S92137/01.}


\abstract{ We give an explicit comparison of eigenvalues and
eigenvectors of $XY$ Hamiltonians of an open linear spin-1/2 chain
and a closed spin-1/2 ring with periodic in space coefficients. It
is shown that the Hamiltonian of a $k$-periodic chain with $nk-1$
sites has $(n-1)k$ multiplicity one eigenvalues which are
eigenvalues of multiplicity two for a Hamiltonian of a
$k$-periodic ring with $2kn$ sites. For the corresponding
eigenvectors in the case of a chain an explicit expression in
terms of eigenvectors of the ring Hamiltonian is given. The
remaining $k-1$ eigenvalues of the chain Hamiltonian and $2k$
eigenvalues of the ring Hamiltonian, together with the
corresponding eigenvectors, are responsible for the difference
between chain and ring models which displays in the boundary
effects at the ends of the chain and translation invariance of the
periodic ring. }


\maketitle

\section{Introduction}

One-dimensional exactly soluble homogeneous spin models (spin
chains, rings) serve as a convenient tool for understanding
quantum dynamics of spin systems~\cite{Gaudin}. Over the past few
years new exactly solvable examples of non-homogeneous spin
systems have been
found~\cite{LimaGon99,FeldmanRud,Feldman,KuzFeld,Ye} and this
prompts a string of applications to the problems of quantum NMR
dynamic~\cite{DorMakFel} and quantum information
theory~\cite{CDEL,Feldman,KuzFeld}. Another possible direction for
the development of the ideas involved in exact solutions for these
models would be deeper understanding of spectra and eigenvectors
of Hamiltonians of non-homogeneous spin systems with different
types of interactions like, for example, Ising
model~\cite{Baxter}. From this point of view it is important to
clarify the difference between open linear spin models and closed
ring spin models.

 Closed ring models usually posses extra symmetries coming, for example, from
translation invariance of the system. Therefore, it is sometimes
easier to study such systems through a framework of quantum
integrals and reduction of the dimension. Elementary physical
arguments show that for a large length a spin ring and a spin
chain behave similar at small times and, therefore, one can expect
to recognize essential chunk of properties for the chain just by
comparing it with the ring.  There are also correction terms which
become significant when the evolution time is long and we have to
take into account those spin wave packets which reflect from the
ends of the chain. Though there is no reason to expect these
correction terms to be universal and independent from particular
properties of the chain, it is still reasonable that their
structure reflects merely the difference between the corresponding
ring and chain models.

 The aim of the present paper is to give a detail comparison of
 spectra and eigenvectors of XY Hamiltonians of an open spin chain
 and a closed spin ring with periodic in space coefficients. The
 simplest example is homogeneous XY models which were solved
 in~\cite{Gon,Lieb}. It corresponds to the case of period one.
 It turns out that all eigenvalues for the chain Hamiltonian in
 homogeneous case have multiplicity one and equal to all but two eigenvalues
 of a homogeneous spin ring with $2N+2$ sites where $N$ is the length of the
open chain. These eigenvalues of the ring Hamiltonian have multiplicity two.
 Corresponding eigenvectors of the chain Hamiltonian form the first
 $N$ components of particularly chosen eigenvectors of the ring Hamiltonian.
Due to the extra symmetry these eigenvectors of the ring Hamiltonian can be
 uniquely reconstructed from the chain eigenvectors. Therefore,
 the difference of these models is only in properties corresponding to the
 remaining two eigenvalues of the ring Hamiltonian. We observe the same behavior
 for every periodic in space chain and ring
 $XY$ models. In particular, $XY$ Hamiltonian of a $k$-periodic spin chain
 with $kn-1$ sites has $k(n-1)$ common eigenvalues with
 $XY$ Hamiltonian of a $k$-periodic spin ring with $2kn$ sites.
 Remaining eigenvalues of these models are responsible for the
 reflection of spin wave packets from the ends of the chain and
 for the translation symmetry of the ring model.

In due course we review some ideas used for the exact solutions of
$XY$ models in various settings, which also allows us to see some
simplifications comparing to the original proofs. We, thus,
present an alternative derivation for the diagonalization of the
$XY$ Hamiltonian of an alternating spin chain with odd number of
sites, first solved in~\cite{FeldmanRud}, and for the
diagonalization of the $XY$ Hamiltonian of any periodic spin ring,
first given in~\cite{AlvesLimaGon}.

\section{The models}

The Hamiltonian of a spin-1/2 open chain with only nearest
neighbor (NN) couplings has the following general form
\begin{equation}
\label{TheorHamCh} H^{chain}=\sum^N_{n=1}\omega_n
I_{nz}+\sum^{N-1}_{n=1} D_{n,n+1} \left(I_{n,x} I_{n+1,x}+I_{n,y}
I_{n+1,y}\right),
\end{equation}
where $\omega_n$, $n=1,...,N$, are the Larmor frequencies,  and
$D_{n,n+1}\neq 0$, $n=1,...,N-1$, are the NN coupling constants.

 The Jordan--Wigner transformation reduces the study of the
Hamiltonian~(\ref{TheorHamCh})  to the diagonalization problem of
\begin{equation}
\label{JWMCh} H^{chain}=2\Omega+D^{chain},
\end{equation}
where
\begin{equation}
\label{GenMatCh} \Omega= \left[\begin{array}{ccccc}
\omega_1& 0&\cdots & 0& 0\\
0& \omega_2&\cdots & 0& 0\\
\vdots&\vdots&\ddots&\vdots&\vdots\\
0& 0&\cdots & \omega_{N-1}& 0\\
0& 0&\cdots & 0& \omega_N\\
\end{array}
\right],\quad D^{chain}= \left[\begin{array}{ccccc}
0& D_1&\cdots & 0& 0\\
D_1& 0&\cdots & 0& 0\\
\vdots&\vdots&\ddots&\vdots&\vdots\\
0& 0&\cdots & 0& D_{N-1}\\
0& 0&\cdots & D_{N-1}& 0\\
\end{array}
\right]
\end{equation}
($D_j=D_{j,j+1}, j=1,\dots,N-1$).

Similarly, the Hamiltonian of a spin-1/2 ring with only nearest
neighbor couplings has the following general form
\begin{equation}
\label{TheorHamR} H^{ring}=\sum^N_{n=1}\omega_n
I_{nz}+\sum^{N}_{n=1} D_{n,n+1} \left(I_{n,x} I_{n+1,x}+I_{n,y}
I_{n+1,y}\right),
\end{equation}
where $\omega_n$, $n=1,...,N$, are the Larmor frequencies,  and
$D_{n,n+1}\neq 0$, $n=1,...,N$, are the NN coupling constants,
and we assume that spin $N+1$ is the same as spin $1$.

 The Jordan--Wigner transformation reduces the study of the
Hamiltonian~(\ref{TheorHamR})  to the diagonalization problem of
\begin{equation}
\label{JWMR} H^{ring}=2\Omega+D^{ring},
\end{equation}
with
\begin{equation}
\label{GenMatR} \Omega= \left[\begin{array}{ccccc}
\omega_1& 0&\cdots & 0& 0\\
0& \omega_2&\cdots & 0& 0\\
\vdots&\vdots&\ddots&\vdots&\vdots\\
0& 0&\cdots & \omega_{N-1}& 0\\
0& 0&\cdots & 0& \omega_N\\
\end{array}
\right],\quad D^{ring}= \left[\begin{array}{ccccc}
0& D_1&\cdots & 0& D_N\\
D_1& 0&\cdots & 0& 0\\
\vdots&\vdots&\ddots&\vdots&\vdots\\
0& 0&\cdots & 0& D_{N-1}\\
D_N& 0&\cdots & D_{N-1}& 0\\
\end{array}
\right],
\end{equation}
where $D_j=D_{j,j+1}$, $j=1,\dots,N$, and the matrix $D^{ring}$
differs from $D^{chain}$ only by the right upper and the left
lower corners.

Assume now that the Larmor frequencies and the coupling constants are
repeating periodically with a period $k$ i.e.,
\begin{equation}
\omega_j=\omega_{j+k},\quad D_j=D_{j+k}.
\end{equation}
In addition to this for the case of the ring model we have to
assume that
$N=km$. We rename the Hamiltonians for the chain and the ring
to underline dependence on $\omega_j$, $D_j$, $j=1,\dots, k$, and
periodicity as
\begin{equation}
H^{chain}_{N}=H^{chain}_{N}(\omega_j,D_j,k);\quad H^{ring}_{km}=
H^{ring}_{km}(\omega_j,D_j,k).
\end{equation}
In what follows it would be convenient for us to consider
independently components of vectors in $\mathbb C^N$ corresponding
to different reminders modulo $k$. We denote these components by
$u_{(j)}$, $j=1,2,\dots k$, so that if
$$
u=(u_1,u_2,\dots,u_N)^t,
$$
then
$$
u_{(j)}=(u_j,u_{j+k},u_{j+2k},\dots)^t.
$$

 For the ring model all components have
exactly $m$ coordinates, while for the chain with $km-1$ sites the
component $u_{(k)}$ has only $m-1$ coordinates and the other
components have exactly $m$ coordinates. We will also fix a
notation
\begin{equation}
H_{i,j} = \left[\begin{array}{ccccc}
2\omega_i& D_i&\cdots & 0& 0\\
D_i& 2\omega_{i+1}&\cdots & 0& 0\\
\vdots&\vdots&\ddots&\vdots&\vdots\\
0& 0&\cdots & 2\omega_{j-1}& D_{j-1}\\
0& 0&\cdots & D_{j-1}& 2\omega_j\\\
\end{array}
\right]
\end{equation}
for a symmetric three-diagonal $(j-i+1)\times (j-i+1)$-matrix with
coefficients $2\omega_i,\dots,2\omega_j$, $D_i,\dots,D_{j-1}$. The
$N\times N$ identity matrix will be denoted by $I_N$.


\section{Homogeneous models}

We begin our comparison by considering the simplest possible case, i.e.
the case of period $k=1$ with equal Larmor frequencies and
coupling constants, which is known as a homogeneous model.
\begin{Th}
\label{HoCha}
$H^{chain}_{N}(\omega,D,1)$ has $N$ distinct eigenvalues
\begin{equation}
\lambda_j=2\omega+2Dcos\left(\frac{\pi j}{N+1}\right),\quad
j=1,\dots,N,
\end{equation}
and the corresponding eigenvectors are of the form
\begin{equation}
\vec u_j=\vec u_j(N)=\left(sin\left(\frac{\pi
j}{N+1}\right),sin\left(\frac{2\pi
j}{N+1}\right),\dots,sin\left(\frac{N\pi j}{N+1}\right)\right).
\end{equation}
\end{Th}
Similarly,
\begin{Th}
\label{HoRi} $H^{ring}_{N}(\omega,D,1)$ has $N-[N/2]$ distinct
eigenvalues
\begin{equation}
\lambda_j=2\omega+2Dcos\left(\frac{2\pi j}{N}\right),\quad 0\le
j\le N/2.
\end{equation}
If $0<j<N/2$ then $\lambda_j$ has multiplicity two and the
corresponding eigenvectors are of the form
\begin{equation}
\vec v_j=\vec v_j(N)=\left(cos\left(\frac{2\pi
j}{N}\right),cos\left(\frac{2\cdot 2\pi
j}{N}\right),\dots,cos\left(\frac{(N-1)\cdot 2\pi
j}{N}\right),1\right),
\end{equation}
\begin{equation}
\vec w_j=\vec w_j(N)=\left(sin\left(\frac{2\pi
j}{N}\right),sin\left(\frac{2\cdot2\pi
j}{N}\right),\dots,sin\left(\frac{2\cdot (N-1)\pi
j}{N}\right),0\right).
\end{equation}
If $j=0$ then $\lambda_j=2\omega+2D$ has multiplicity one with an
eigenvector
\begin{equation}
\vec u=(1,1,\dots,1).
\end{equation}
If $N$  is even and $j=N/2$ then $\lambda_j=2\omega-2D$ is an
eigenvalue of multiplicity one with an eigenvector
\begin{equation}
\vec u=(1,-1,1,-1,\dots,1,-1).
\end{equation}
\end{Th}
\begin{Rem}
Theorems~\ref{HoCha},~\ref{HoRi} were obtained in~\cite{Gon,Lieb}.
Proofs of both the theorems are straight forward verifications.
\end{Rem}
\begin{Cor}
\label{HomCR} The spectra of $H^{chain}_N(\omega,D,1)$ and
$H^{ring}_{2N+2}(\omega,D,1)$ are related to each other by
$$
det\left(H^{chain}_N(\omega,D,1)-\lambda
I_N\right)^2\left(2\omega+2D-\lambda\right)
\left(2\omega-2D-\lambda\right) =
$$
\begin{equation}
=
det \left( H^{ring}_{2N+2}(\omega,D,1)-\lambda I_{2N+2}\right).
\end{equation}
For $j=1,\dots,N$ the following relation between corresponding
eigenvectors for the chain and the ring models holds:
\begin{equation}
\vec u_j=\Pi^{2N+2}_{N}\vec w_j,
\end{equation}
where for $M>L$
\begin{equation}
\Pi^M_L:\mathbb R^M\to \mathbb R^L
\end{equation}
is the projection onto the first $L$-coordinates.
\end{Cor}


\section{Alternating models}

In this section we compare alternating spin-1/2 chain and ring
models, i.e. models of period 2. We will discuss only open spin
chains with odd number of sites as the structure of this model in
the even case contains transcendental equation~\cite{KuzFeld}
which does not feature in the ring model (see below).

To diagonalize the Hamiltonian of an open alternating linear
spin-1/2 chain with odd number of sites we shall rewrite the
initial three-diagonal matrix for the Hamiltonian
$H^{chain}_{2n-1}(\omega_1,\omega_2,D_1,D_2,2)$ in a different basis
and represent it
as a block
$2\times 2$ matrix with respect to odd $u_{(1)}$ and even $u_{(2)}$
components of vectors:
\begin{equation}
H^{chain}_{2n-1}(\omega_1,\omega_2,D_1,D_2,2)
\left(\begin{array}{c}
u_{(1)}\\
u_{(2)}\\
\end{array}\right)
=
\left[\begin{array}{cc}
2\omega_1 I_{n}& {\cal L}^t_{chain}\\
{\cal L}_{chain}& 2\omega_2 I_{n-1}\\
\end{array}\right]
\left(\begin{array}{c}
u_{(1)}\\
u_{(2)}\\
\end{array}\right)
\end{equation}
where ${\cal L}_{chain}: {\mathbb R}^{n}\to {\mathbb R}^{n-1}$:
\begin{equation}
{\cal L}_{chain}= \left[\begin{array}{ccccccc}
D_1    &D_2    &0   &\cdots& 0&0&0\\
0      &D_1    &D_2 &\cdots& 0&0&0\\
0      &0      &D_1 &\ddots& 0&0&0\\
\vdots &\vdots &\ddots&\ddots&\ddots&\vdots&\vdots\\
0      &0      &0&\ddots&\ddots&\ddots&\vdots\\
0      &0      &0&\cdots&0 &\ddots&\ddots\\
\end{array}\right].
\end{equation}
Observe that $2\omega_1$ is an eigenvalue of
$H^{chain}_{2n-1}(\omega_1,\omega_2,D_1,D_2,2)$ of multiplicity
one with an eigenvector $u$ having the following components:
\begin{equation}
u_{(1)}\in Ker{\cal L}_{chain},\quad u_{(2)}=0.
\end{equation}
If $\lambda\neq 2\omega_1$ is an eigenvalue for
$H^{chain}_{2n-1}(\omega_1,\omega_2,D_1,D_2,2)$ then
\begin{equation}
u_{(1)}=\frac{1}{\lambda-2\omega_1}{\cal L}^t_{chain}
u_{(2)},\quad (\lambda-2\omega_2)u_{(2)}={\cal L}_{chain} u_{(1)}.
\end{equation}
This implies that
\begin{equation}
(\lambda-2\omega_1)(\lambda-2\omega_2)u_{(2)}={\cal
L}_{chain}{\cal L}^t_{chain} u_{(2)}.
\end{equation}
${\cal L}_{chain}{\cal L}^t_{chain}$ is a tridiagonal matrix with
all elements on the main diagonal equal to $D^2_1+D^2_2$ and all
off-diagonal elements equal to $D_1D_2$. Using explicit
diagonalization of the Hamiltonian of a homogeneous chain from the
previous section we deduce
\begin{Th}
\label{AltCh} Each eigenvalue of the Hamiltonian
$H^{chain}_{2n-1}(\omega_1,\omega_2,D_1,D_2,2)$ of an alternating
spin system with odd number of sites is either $2\omega_1$ or it
is a solution of the equation
\begin{equation}
\label{AltOp} \det\left(H_{1,2}-\lambda I_2\right)
-D^2_2=2D_1D_2cos\left(\frac{\pi j}{n}\right),\quad
j=1,2,\dots,n-1.
\end{equation}
All eigenvalues have multiplicity one and if $\lambda$ is a root
of~(\ref{AltOp}) for some $j=1,2,\dots, n-1$, then it is an
eigenvalue for $H^{chain}_{2n-1}(\omega_1,\omega_2,D_1,D_2,2)$ and
the component $u_{(2)}$ of the corresponding eigenvector
$u_{\lambda}$ is
\begin{equation}
u_{(2)}=\left(sin\left(\frac{\pi
j}{n}\right),\dots,sin\left(\frac{(n-1)\pi j}{n}\right)\right).
\end{equation}
The other component $u_{(1)}$ is given by:
\begin{equation}
u_{(1)}=\frac{1}{\lambda-2\omega_1}{\cal L}^t_{chain}u_{(2)}.
\end{equation}
An eigenvector corresponding to $2\omega_1$ has $u_{(2)}=0$ and
$u_{(1)}$ spans the one-dimensional kernel of ${\cal L}_{chain}$.
\end{Th}
\begin{Rem}
Theorem~\ref{AltCh} was discovered in~\cite{FeldmanRud}. The case
of even number of sites in open alternating spin-1/2 chain models
was solved in~\cite{KuzFeld}. Arguments given before the statement
of this theorem yield another proof of the result
from~\cite{FeldmanRud}.
\end{Rem}

\

\noindent
Now we consider the ring case. Set $q_l=exp\left(\frac{2\pi
l}{n}\right)$, $l=0,1,\dots,n-1$ and
\begin{equation}
U_l=(1,0,q_l,0,q^2_l,\dots,q^{n-1}_l,0),\quad
V_l=(0,1,0,q_l,0,\dots,0,q^{n-1}_l),
\end{equation}
then
\begin{equation}
H^{ring}_{2n}(\omega_1,\omega_2,D_1,D_2,2)U_l=2\omega_1U_l+\left(D_1+D_2q_l\right)
V_l,
\end{equation}
\begin{equation}
H^{ring}_{2n}(\omega_1,\omega_2,D_1,D_2,2)V_l=\left(D_2q^{-1}_l+D_1\right)U_l+2\omega_2V_l.
\end{equation}
Therefore, $\mu U_l+\kappa V_l$ is an eigenvector for
$H^{ring}_{2n}$ with eigenvalue $\lambda$ if $(\mu,\kappa)$ is an
eigenvector for
\begin{equation}
\left(\begin{array}{cc}
2\omega_1&D_2q^{-1}_l+D_1\\
D_1+D_2q_l&2\omega_2\\
\end{array}
\right)
\end{equation}
with eigenvalue $\lambda$. Because,
\begin{equation}
\left(D_2q^{-1}_l+D_1\right)\left(D_1+D_2q_l\right)=D^2_1+D^2_2+2D_1D_2cos\left(\frac{2\pi
l}{n}\right)>0
\end{equation}
we conclude that $2\omega_1$ is not an eigenvalue for
$H^{ring}_{2n}$ and
\begin{equation}
\mu=\frac{D_2q^{-1}_l+D_1}{\lambda-2\omega_1}\kappa.
\end{equation}
This leads to the following theorem.
\begin{Th}
\label{AltRin}
Each eigenvalue of the Hamiltonian
$H^{ring}_{2n}(\omega_1,\omega_2,D_1,D_2,2)$ is a solution to the
equation
\begin{equation}
\label{AltRi}
 \det\left(H_{1,2}-\lambda
I_2\right)-D^2_2=2D_1D_2cos\left(\frac{2\pi j}{n}\right)
\end{equation}
for some $0\le j\le n/2$. If $\lambda_j$ is a solution
to~(\ref{AltRi}) with $0<j<n/2$ then it is an eigenvalue of
$H^{ring}_{2n}$ of multiplicity two. Components $u_{(2)}$
of the corresponding eigenvector $u_{\lambda_j}$
can be
chosen as
\begin{equation}
\label{COS}
\left(cos\left(\frac{2\pi j}{n}\right),cos\left(\frac{2\cdot
2\pi j}{n}\right),\dots,cos\left(\frac{(n-1)\cdot 2\pi
j}{n}\right),1\right),
\end{equation}
or
\begin{equation}
\label{SIN}
\left(sin\left(\frac{2\pi j}{n}\right),sin\left(\frac{2\cdot
2\pi j}{n}\right),\dots,sin\left(\frac{(n-1)\cdot 2\pi
j}{n}\right),0\right).
\end{equation}
If $\lambda_j$ is a solution to~(\ref{AltRi}) with $j=0$ or $j=n/2$
then it is an eigenvalue of $H^{ring}_{2n}$ of multiplicity one.
For $j=0$ the component $u_{(2)}$ can be chosen as
\begin{equation}
u_{(2)}=(1,1,1,\dots,1,1),
\end{equation}
and for $j=n/2$
\begin{equation}
u_{(2)}=(1,-1,1,-1,\dots,1,-1)
\end{equation}

In all cases
\begin{equation}
u_{(1)}=\frac{1}{\lambda-2\omega_1}{\cal L}^t_{ring}u_{(2)},
\end{equation}
where $n\times n$ matrix ${\cal L}_{ring}$ is given by:
\begin{equation}
{\cal L}_{ring}= \left[\begin{array}{cccccc}
D_1    &D_2    &0   &\cdots& 0&0\\
0      &D_1    &D_2 &\cdots& 0&0\\
0      &0      &D_1 &\ddots& 0&0\\
\vdots &\vdots &\ddots&\ddots&\vdots&\vdots\\
0      &0      &0&\ddots&\ddots&\vdots\\
D_2      &0      &0&\cdots &0&D_1\\
\end{array}\right].
\end{equation}
\end{Th}
\begin{Rem}
Theorem~\ref{AltRin} was proved in~\cite{LimaGon99} and~\cite{Ye}.
Arguments given before the statement of the theorem comprise an
alternative proof of this result.
\end{Rem}
\begin{Cor}
\label{AltCR} The spectra of
$H^{chain}_{2n-1}(\omega_1,\omega_2,D_1,D_2,2)$ and
$H^{ring}_{4n}(\omega_1,\omega_2,D_1,D_2,2)$ are related to each
other by means of
$$
\det\left(H^{chain}_{2n-1}(\omega_1,\omega_2,D_1,D_2,2)-\lambda I_{2n-1}\right)^2
\left(\det(H_{1,2}-\lambda I_2)-D^2_2-2D_1D_2\right)\times
$$
$$
\times
\left(\det(H_{1,2}-\lambda I_2)-D^2_2+2D_1D_2\right)=
$$
\begin{equation}
=\det\left(H^{ring}_{4n}(\omega_1,\omega_2,D_1,D_2,2)-\lambda I_{4n}\right)
(2\omega_1-\lambda)^2
\end{equation}
If $u_j$ and $v_j$ are eigenvectors of
$H^{chain}_{2n-1}(\omega_1,\omega_2,D_1,D_2,2)$ and
$H^{ring}_{4n}(\omega_1,\omega_2,D_1,D_2,2)$ respectively,
corresponding to the same eigenvalue $\lambda_j$,
$j=1,2,\dots,n-1$, and $v_j$ is of the form~(\ref{SIN}) then
\begin{equation}
u_j=\Pi^{4n}_{2n-1} v_j.
\end{equation}
\end{Cor}
\pf  The first part of the theorem comes from comparison
of the spectra for $H^{chain}_{2n-1}$ and $H^{ring}_{4n}$ given in
Theorems~\ref{AltCh},~\ref{AltRin}. The second part is a
consequence of the structure of the component $u_{(2)}$ of the
corresponding eigenvectors (in particular, that the $n$-th
coordinate of $u_{(2)}$ in the ring case is zero) and the fact
that ${\cal L}_{chain}$ forms the $(n-1)\times n$-left upper
corner of ${\cal L}_{ring}$.
\eop


\section{Diagonalization of an XY Hamiltonian of spin-1/2 rings with periodic coupling constants and
Larmor frequencies}

We now proceed with the most general case of periodic models and
give an alternative to~\cite{AlvesLimaGon} diagonalization
procedure for a Hamiltonian of a spin-1/2 ring with periodic
coefficients of any period. Observe some elementary properties of
quantum integrals for such systems. Let $T_N:\mathbb R^N\to
\mathbb R^N$ be
\begin{equation}
\label{GenMat} T_N= \left[\begin{array}{cccccc}
0& 1& 0&\cdots & 0& 0\\
0& 0& 1&\cdots & 0& 0\\
\vdots&\vdots& \vdots&\ddots&\vdots&\vdots\\
0& 0& 0&\cdots & 0& 1\\
1& 0& 0&\cdots & 0& 0\\
\end{array}
\right].
\end{equation}
Performing explicitly matrix multiplication we deduce that
$$
T_{km}
H^{ring}_{km}(\omega_1,\dots,\omega_{k-1},\omega_k,D_1,\dots,D_{k-1},
D_k)T^t_{km}=
$$
\begin{equation}
=
H^{ring}_{km}(\omega_2,\dots,\omega_k,\omega_1,D_2,\dots,D_k,D_1).
\end{equation}
In particular,
\begin{equation}
\label{ComRel} \left(T_{km}\right)^k H^{ring}_{km}=H^{ring}_{km}
\left(T_{km}\right)^k.
\end{equation}
Consider two collections of roots of one
\begin{equation}
p_j=e^{\frac{2\pi ij}{k}},\quad j=0,1,\dots,k-1,\quad
r_l=e^{\frac{2\pi il}{km}},\quad l=0,1,\dots,m-1,\quad
\end{equation}
and for each pair $(j,l)$, $j=0,1,\dots,k-1$, $l=0,1,\dots,m-1$,
let us introduce a vector
\begin{equation}
V_{jl}=\left(1,p_jr_l,p^2_jr^2_l,\dots,p^{km-1}_jr^{km-1}_l\right)^t.
\end{equation}
Obviously,
\begin{equation}
T_{km} \left(V_{jl}\right)= p_jr_l
V_{jl},
\end{equation}
and, therefore, from commutation relation~(\ref{ComRel})
$\tilde H_{km}\left(V_{jl}\right)$ belongs to a
subspace spanned by $V_{0,l},\dots,V_{k-1,l}$.
\begin{Lemm}
\label{SUBSP}
A subspace of $\mathbb R^{km}$ spanned by $V_{0,l},\dots,V_{k-1,l}$
coincides with a subspace spanned by
\begin{equation}
V_{l},\quad \left(T^t_{km}\right)\left(\tilde V_{l}\right),\quad
\dots\quad
\left(T^t_{km}\right)^{k-1}\left(\tilde V_{l}\right),
\end{equation}
where
\begin{equation}
\tilde
V_{l}=(1,\underbrace{0,\dots,0}_{k-1},q_l,\underbrace{0,\dots,0}_{k-1},q^{2}_l,0,\dots,0,q^{k-1}_l,
\underbrace{0,\dots,0}_{k-1})^t,\quad q_l=e^{\frac{2\pi il}{m}}
\end{equation}
\end{Lemm}
\pf Let us introduce vectors
\begin{equation}
U_j=\left(1,p_j,p^2_j,\dots,p^{k-1}_j\right)^t,\quad j=0,1,\dots,k-1,
\end{equation}
and the corresponding Vandermonde $k\times k$-matrix
\begin{equation}
W_k=\left( U_0,U_1,\dots,U_{k-1}\right).
\end{equation}
Performing explicitly matrix multiplication we deduce that
\begin{equation}
\left( V_{0,l},\dots,V_{k-1,l}\right) W^{-1}_k= \left(\tilde
V_{l},r_l\left(T^t_{km}\right)\left(\tilde V_{l}\right),
\dots,r^{k-1}_l\left(T^t_{km}\right)^{k-1}\left(\tilde
V_{l}\right)\right),
\end{equation}
which implies the statement.
\eop

\noindent
For every $l=0,\dots,m-1$ we define
\begin{equation}
H_k(q_l)= \left[\begin{array}{cccccc}
2\omega_1& D_1&0&\cdots & 0& q^{-1}_lD_k\\
D_1& 2\omega_2& D_2&\cdots & 0& 0\\
0&D_2&2\omega_3&\cdots &0&0\\
\vdots&\vdots&\vdots& \ddots&\vdots&\vdots\\
0& 0&0& \cdots & 2\omega_{k-1}& D_{k-1}\\
q_lD_k& 0&0&\cdots & D_{k-1}& 2\omega_k\\
\end{array}
\right].
\end{equation}
\begin{Th}
\label{PerRin} Each eigenvalue of $H^{ring}_{km}(\omega_j,D_j,k)$
is an eigenvalue of $H_k(q_l)$, for some $l=1,\dots,m$. If
$\left(\mu_1,\mu_2,\dots,\mu_{k}\right)^t$ is an an eigenvector of
$H_k(q_l)$ then
\begin{equation}
\label{EiVeStr}
\mu_1 \tilde V_{l}+\mu_2 \left(T^t_{km}\right)\left(\tilde
V_{l}\right)+\cdots +\mu_{k}
\left(T^t_{km}\right)^{k-1}\left(\tilde V_{l}\right),
\end{equation}
with
\begin{equation}
\tilde
V_{l}=(1,\underbrace{0,\dots,0}_{k-1},q_l,\underbrace{0,\dots,0}_{k-1},q^{2}_l,0,\dots,0,q^{k-1}_l,
\underbrace{0,\dots,0}_{k-1})^t,
\end{equation}
is an eigenvector of $H^{ring}_{km}(\omega_j,D_j,k)$.
\end{Th}
\pf  Note that ${\cal V}_l= \langle V_{0l}, V_{1l},\dots
V_{k-1,l} \rangle$ is invariant under $H^{ring}_{km}$ and, thus,
we can work with each subspace ${\cal V}_l$ independently. Let
$V\in {\cal V}_l$ be an eigenvector for $H^{ring}_{km}$ with
eigenvalue $\kappa$. From Lemma~\ref{SUBSP} we can assume that
\begin{equation}
V=\mu_1\tilde V_{l}+\mu_2 T^t_{km}\left(\tilde
V_{l}\right)+\cdots+
\mu_{k}\left(T^t_{km}\right)^{k-1}\left(\tilde V_{l}\right).
\end{equation}
We shall verify that $(\mu_1,\mu_2,\dots,\mu_k)$ is an eigenvector
for $H_k(q_l)$ with the same eigenvalue $\kappa$. Indeed, look at
the $(ks+t)$ row of $H^{ring}_{km}$, $0\le s\le m-1$, $1\le t\le
k$. If $1<t<k$ then multiplying it with $V$ we get
\begin{equation}
\mu_{t-1} q^{s}_lD_{t-1} +\mu_t q^{s}_l
2\omega_t+\mu_{t+1}q^{s}_lD_t.
\end{equation}
For $t=1$ and $t=k$ we obtain respectively:
\begin{equation}
\mu_{k} q^{s-1}_lD_{k}+\mu_{1} q^s_l 2\omega_{1}+\mu_2 q^{s}_l
D_2,
\end{equation}
and
\begin{equation}
\mu_{k-1} q^{s}_lD_{k-1}+\mu_k q^{s}_l 2\omega_k+\mu_1 q^{s+1}_l
D_k.
\end{equation}
Comparing it with the $ks+t$ coordinate of $\kappa V$ and dividing
both sides by $q^{s}_l$ we get exactly the $t$-th row of
\begin{equation}
H_k(q_l)(\mu_1,\dots,\mu_k)^t=\kappa(\mu_1,\dots,\mu_k)^t,
\end{equation}
which is what we need.
\eop
\begin{Rem}
Theorem~\ref{PerRin} was proved in~\cite{AlvesLimaGon}. The proof
of the result given above presents different approach compared
to~\cite{AlvesLimaGon} where the authors used Fourier transform.
\end{Rem}
We finish this section with one useful lemma.
\begin{Lemm}
\label{QuaEq} The characteristic polynomial of $H_k(q_l)$ can be
rewritten as:
$$
det(H_k(q_l)-\lambda I_k)=
$$
\begin{equation}
=
det(H_{1,k}-\lambda I_k)-
det(H_{2,k-1}-\lambda I_{k-2})D^2_k -(-1)^k2D_1\cdots D_k
cos\left(\frac{2\pi  l}{m}\right)
\end{equation}
\end{Lemm}
\pf Consider $det(H_k(q_l)-\lambda I_k)$ as a polynomial
in $D_k$. It is a quadratic polynomial with free term
\begin{equation}
det(H_{1,k}-\lambda I_k).
\end{equation}
The coefficient of $D^2_k$ is
\begin{equation}
(-1)^{k+1+k-1+1}det(H_{2,k-1}-\lambda I_{k-2})
\end{equation}
and, finally, the coefficient of $D_k$ is
\begin{equation}
(-1)^{k+1}D_1\cdots D_{k-1} \left(q_l+q^{-1}_l\right)
=-(-1)^k2D_1\cdots D_{k-1} cos\left(\frac{2\pi  l}{m}\right).
\end{equation}
\eop


\section{Comparison}

In this section we give a comparison of periodic chain and ring
models similar to Corollaries~\ref{HomCR},~\ref{AltCR} for any
period $k$. Let ${\cal Q}_{chain},{\cal R}_{chain}:{\mathbb
R}^{n}\to {\mathbb R}^{n-1}$ be given by
\begin{equation}
\label{DdDk} {\cal Q}_{chain}=\left[\begin{array}{ccccccc}
1    &0    &0 &\cdots& 0&0&0\\
0    &1    &0 &\cdots& 0&0&0\\
0    &0    &1 &\ddots& 0&0&0\\
\vdots &\vdots &\vdots&\ddots&\ddots&\vdots&\vdots\\
0&0&0&\cdots&1&0 &0\\
0&0&0&\cdots&0 &1&0\\
\end{array}\right],\quad
{\cal R}_{chain}=\left[\begin{array}{ccccccc}
0     &1  &0 &\cdots& 0&0&0\\
0     &0    &1 &\cdots& 0&0&0\\
0     &0    &0 &\ddots& 0&0&0\\
\vdots&\vdots &\vdots&\ddots&\ddots&\vdots&\vdots\\
0&0&0&\cdots&0 &1 &0\\
0&0&0&\cdots&0 &0   &1\\
\end{array}\right].
\end{equation}
\begin{Th}
\label{MAINkn-1} Each eigenvalue of the Hamiltonian
$H^{chain}_{kn-1}(\omega_j,D_j,k)$ of a $k$-periodic system with
$kn-1$ sites is either an eigenvalue of $H_{1,k-1}$ or it is a
solution of the equation
\begin{equation}
\label{EQU} det\left(H_{1,k}-\lambda I_k\right)
-det\left(H_{2,k-1}-\lambda I_{k-2}\right)D^2_k= (-1)^k
2D_1\cdots D_k cos\left(\frac{\pi j}{n}\right),
\end{equation}
for some $j=1,...,n-1$. Equation~(\ref{EQU}) does not have
repeated roots and all $k(n-1)$ solutions constructed
from~(\ref{EQU}) are pairwise distinct and are not eigenvalues of
$H_{1,k-1}$.

If $\lambda $ is the solution of~(\ref{EQU}) for some
$j=1,...,n-1$, then it is an eigenvalue of
$H^{chain}_{kn-1}(\omega_j,D_j,k)$ and the component $u_{(k)}$ of
the corresponding eigenvector $u_{\lambda}$ is
\begin{equation}
u_{(k)}=\left(sin\left(\frac{\pi j}{n}\right),\dots,
sin\left(\frac{(n-1)\pi j}{n}\right)\right).
\end{equation}
The other components $u_{(j)}$, $j=1,...,k-1$, are determined
uniquely from
$$
u_{(j)}=\frac{(-1)^{j}}{det\left(H_{1,k-1}-\lambda
I_{k-1}\right)}\cdot \left[D_1\cdots D_{j-1}
det\left(H_{j+1,k-1}-\lambda I_{k-j-1}\right)\cdot D_k{\cal
R}^t_{chain}+ \right.
$$
\begin{equation}
\label{QRP} \left. +(-1)^{k}det\left(H_{1,j-1}-\lambda
I_{j-1}\right) D_j\cdots D_{k-2}D_{k-1}{\cal
Q}^t_{chain}\right]u_{(k)},
\end{equation}
where ${\cal Q}_{chain}$, ${\cal R}_{chain}$ are given
in~(\ref{DdDk}). Every eigenvalue $\lambda $ of $H_{1,k-1}$
is an eigenvalue of $H$. The component $u_{(k)}$ of the
corresponding eigenvector  of $H^{chain}_{kn-1}(\omega_j,D_j,k)$
is zero. The component
$u_{(1)}$ spans the one-dimensional kernel of
\begin{equation}
(-1)^{k-1}D_1\cdots D_{k-2}D_{k-1}{\cal Q}_{chain}-
det\left(H_{2,k-1}-\lambda I_{k-2}\right)D_k{\cal R}_{chain}.
\end{equation}
The remaining components $u_{(j)}$, $j=2,...,k-1$, are
\begin{equation}
\label{THEOTHER} u_{(j)}=(-1)^{j-1}\frac{D_1\cdots D_{j-1}
det\left(H_{j+1,k-1}-\lambda I_{k-j-1}\right)}{det\left(H_{2,k-1}-\lambda
I_{k-2}\right)}u_{(1)}.
\end{equation}
\end{Th}
\begin{Rem}
A complete proof of this statement is given in~\cite{Feldman}.
\end{Rem}
\begin{Th}
\label{MAINRing}
 Each eigenvalue of $H^{ring}_{km}(\omega_j,D_j,k)$ is a solution of the equation
\begin{equation}
\label{EQ}
 det(H_{1,k}-\lambda I_k)- det(H_{2,k-1}-\lambda
I_{k-2})D^2_k =(-1)^k2D_1\cdots D_k cos\left(\frac{2\pi
l}{m}\right)
\end{equation}
for some $0\le l\le m/2$. If $\lambda_l$ is a solution
of~(\ref{EQ}) for some $0<l<m/2$, then it is an eigenvalue of
$H^{ring}_{km}(\omega_j,D_j,k)$ of multiplicity 2 and the
component $u_{(k)}$ of the corresponding eigenvector $u_l$ is
either
\begin{equation}
\label{V1} \left(cos\left(\frac{2\pi
l}{m}\right),cos\left(\frac{2\cdot 2\pi
l}{m}\right),\dots,cos\left(\frac{(m-1)2\pi l}{m}\right),1\right),
\end{equation}
or
\begin{equation}
\label{V2}
\left(sin\left(\frac{2\pi
l}{m}\right),sin\left(\frac{2\cdot 2\pi
l}{m}\right),\dots,sin\left(\frac{(m-1)2\pi l}{m}\right),0\right).
\end{equation}
If $\lambda_l$ is a solution to~(\ref{AltRi}) with $l=0$ or
$l=m/2$ then it is an eigenvalue of $H^{ring}_{km}$ of
multiplicity one. For $l=0$ the component $u_{(k)}$ can be chosen
as
\begin{equation}
u_{(k)}=(1,1,1,\dots,1,1),
\end{equation}
and for $l=m/2$
\begin{equation}
u_{(k)}=(1,-1,1,-1,\dots,1,-1).
\end{equation}
In all cases the other components $u_{(j)}$, $j=1,\dots,k-1$, are
determined uniquely from
$$
u_{(j)}=\frac{(-1)^{j}}{det\left(H_{1,k-1}-\lambda
I_{k-1}\right)}\left[D_1\dots D_{j-1} det\left(H_{j+1,k-1}-\lambda
I_{k-j-1}\right)D_kT^t_m\right.+
$$
\begin{equation}
\label{VectFi} + \left.(-1)^{k}det\left(H_{1,j-1}-\lambda
I_{j-1}\right) D_j\cdots D_{k-2}D_{k-1}\right]u_{(k)},
\end{equation}
where $T_m$ is given by~(\ref{GenMat}) ($N=m$).
\end{Th}
\pf  The eigenvalue part of the theorem  is a consequence
of Theorem~\ref{PerRin} and Lemma~\ref{QuaEq}. To understand the
structure of eigenvectors we consider the matrix
$P=\left(H_{1,k-1}-\lambda I_{k-1}\right)^{-1}$ and denote its
elements by $P_{i,j}$ (where $i$ is row, and $j$ is column,
$i,j=1,\dots k-1$). The following explicit expressions were
deduced in~\cite{Feldman} for $t>s$
\begin{equation}
\label{E1}
P_{t,s}=P_{s,t}=(-1)^{s+t}\frac{det\left(H_{1,s-1}-\lambda I_{s-1}\right)
D_{s}\cdots D_{t-1}det\left(H_{t+1,k-1}-\lambda I_{k-t-1}\right)}
{det\left(H_{1,k-1}-\lambda I_{k-1}\right)},
\end{equation}
and
\begin{equation}
\label{E2}
P_{t,t}=\frac{det\left(H_{1,t-1}-\lambda I_{t-1}\right)
det\left(H_{t+1,k-1}-\lambda I_{k-t-1}\right)}
{det\left(H_{1,k-1}-\lambda I_{k-1}\right)}.
\end{equation}
If $\vec
v=(\mu_1,\dots,\mu_k)$ is an eigenvector for $H_k(q_l)$ with
eigenvalue $\lambda$ then for the matrix
\begin{equation}
G=\left(\begin{array}{cc}
P& 0\\
0& 1\\
\end{array}
\right)
\end{equation}
we have
\begin{equation}
G\cdot (H_k(q_l)-\lambda I_k) \vec v=0,
\end{equation}
or in another form
\begin{equation}
\left(
\begin{array}{cccccc}
1 &  0&\cdots & 0& 0\\
0 &  1&\cdots & 0& 0\\
\vdots & \ddots&\ddots&\ddots& \vdots\\
0 &  0& \cdots & 1& 0\\
0 &  0& \cdots & 0& 1\\
D_kq_l &  0& \cdots & 0& D_{k-1}\\
\end{array}
G\left(\begin{array}{c}
D_kq^{-1}_l\\
0\\
\vdots\\
0\\
D_{k-1}\\
2\omega_k-\lambda\\
\end{array}\right)\right)
\left(\begin{array}{c}
\mu_1\\
\mu_2\\
\vdots\\
\mu_{k-2}\\
\mu_{k-1}\\
\mu_{k}\\
\end{array}\right)=0.
\end{equation}
This implies that
\begin{equation}
\mu_j=-\left(P_{j,1}D_kq^{-1}_l+P_{j,k-1}D_{k-1}\right)\mu_k.
\end{equation}
Substituting this into~(\ref{EiVeStr}) and taking into
account~(\ref{E1}),~(\ref{E2}) we obtain~(\ref{VectFi}).
\eop
\begin{Cor}
The spectra of $H^{chain}_{km-1}(\omega_j,D_j,k)$ and
$H^{ring}_{2km}(\omega_j,D_j,k)$ are related to each other by
means of
$$
\det\left(H^{chain}_{km-1}(\omega_j,D_j,k)-\lambda
I_{km-1}\right)^2 \times
$$
$$
\times \left(\det(H_{1,k}-\lambda I_k)-\det(H_{2,k-1}-\lambda
I_{k-2})D^2_k -2D_1\cdots D_k\right)\times
$$
$$
\times \left(\det(H_{1,k}-\lambda I_k)-\det(H_{2,k-1}-\lambda
I_{k-2})D^2_k +2D_1\cdots D_k\right) =
$$
\begin{equation}
= \det\left(H^{ring}_{2km}(\omega_j,D_j,k)-\lambda I_{2km}\right)
\det(H_{1,k-1}-\lambda I_{k-1})^2
\end{equation}
If $u_j$ and $v_j$ are eigenvectors of
$H^{chain}_{km-1}(\omega_j,D_j,k)$ and
$H^{ring}_{2km}(\omega_j,D_j,k)$ respectively, corresponding to
the same eigenvalue $\lambda_j$, $j=1,2,\dots,m-1$, and $v_j$ is
of the form~(\ref{V2}) then
\begin{equation}
u_j=\Pi^{2km}_{km-1} v_j.
\end{equation}
\end{Cor}
\pf The proof repeats the proof of Corollary~\ref{AltCR}.
The eigenvalue part is a direct consequence of the structure of
the spectra from Theorems~\ref{MAINkn-1}, \ref{MAINRing}. The
eigenvector part is a consequence of the formulae~(\ref{QRP}) and
(\ref{VectFi}), because matrix ${\cal Q}_{chain}$ is a left upper
corner of $I_{2m}$, ${\cal R}_{chain}$ is a left upper corner of
$T_m$, $u_{(k)}$ from Theorem~\ref{MAINkn-1} forms the first $m-1$
coordinates of $u_{(k)}$ from~(\ref{V2}), and the $m$-th
coordinate of $u_{(k)}$ from~(\ref{V2}) is zero.
\eop

\section{Conclusion}

We gave an explicit comparison of the spectrum properties of the
$XY$ Hamiltonians of open spin chain and closed spin ring models
with periodic in space coefficients and identified the part of
spectra responsible for the reflection of spin wave packets from
the ends of the chain and for the translation symmetry of the
ring. Common parts of the spectra for $k$-periodic chain with
$kn-1$ sites and $k$-periodic ring with $2kn$ sites correspond to
the same evolution of these systems at short times.

One can argue that a similar type of behavior must be featured in
other one-dimensional spin systems including the one-dimensional
Ising model in the transverse magnetic field. In particular, using
a recent solution of a periodic Ising model on a ring
in~\cite{Gon07} we might be able to recover an exact
diagonalization for the Hamiltonian of the corresponding model on
an open chain. Experimental results~\cite{Khitrin} yield another
perspective of the development of these analytical findings. Some
problems of quantum information theory (for example, boundary
effects in the study of entanglement in one-dimensional
systems~\cite{Wang}, or relations between entanglement and qubit
addressing~\cite{PyrFel}) provide new possible applications of the
methods suggested.

\end{document}